\documentclass[aps,prl,showpacs,twocolumn]{revtex4}

\usepackage{amsmath}
\usepackage{graphicx}
\usepackage{dcolumn}
\usepackage{bm}

\begin{document}

{\bf Comment on ``Multiple Bosonic Mode Coupling in the Electron Self-Energy of (La$_{2-x}$Sr$_x$)CuO$_4$"}\\

Recently Zhou \textit{et al} reported photoemission data from underdoped
(La$_{2-x}$Sr$_x$)CuO$_4$ revealing "fine structure" in the single-particle 
self-energy $\Sigma(\omega)$ \cite{Zhou}. Four "fine structures" 
at energies of (40\textendash 46) meV, (58\textendash 63) meV, (23\textendash 29) meV and (75\textendash 85) meV, were identified
in the real part of $\Sigma(\omega)$ and attributed to coupling of electrons to four phonon modes. 
The maximum entropy method was used to fit the measured Re$\Sigma(\omega)$ 
and the coupling function $\alpha^2F$ was extracted. Here, we argue that the features in Re$\Sigma(\omega)$
interpreted by Zhou \textit{et al} \cite{Zhou} as the "fine structure" could not be detected with the experimental parameters used in \cite{Zhou}. 
We show that the measured Re$\Sigma(\omega)$ displays more "structure" than physically possible 
and the "fine structure" should therefore be interpreted as statistical noise or other experimental error.
 
Our strategy is to check whether such fine structure may be observed, even in principle, 
in an experiment with the experimental resolution $\Delta_{exp}$ comparable to the splitting between  
neighboring peaks or shoulders ("structures") in the Re$\Sigma(\omega)$. 
As a first step, we took $\alpha^2F$ from \cite{Zhou} and reproduced 
the calculated self-energy, as shown in Fig 1c). This self-energy was directly compared to the experimental Re$\Sigma(\omega)$ in \cite{Zhou}, 
clearly demonstrating that no experimental broadening was considered in the fitting procedure. However, the claimed energy resolution 
in \cite{Zhou} ranges from 12 to 20 meV \cite{res} at which the effects on the measured Re$\Sigma(\omega)$ are quite significant and, as we show bellow,
essentially forbid the observation of "fine structure" suggested in \cite{Zhou}.

The measured Re$\Sigma(\omega)$ is extracted from momentum 
distribution curves (MDC) and is affected by the experimental 
resolution in a non-trivial way. To see the effects of finite energy resolution on the MDC derived
self-energy, we have simulated the photoemission intensity by using 
the self-energy and the dispersion relation from \cite{Zhou}. A constant term of 50 meV (roughly one half of the experimental value)
is added to the Im$\Sigma(\omega)$. The resulting simulation is 
shown in Fig. 1a. Now, the "experimental" spectrum would be smeared in energy and momentum 
by finite energy and momentum resolution. 
\begin{figure}
\begin{center}
\includegraphics {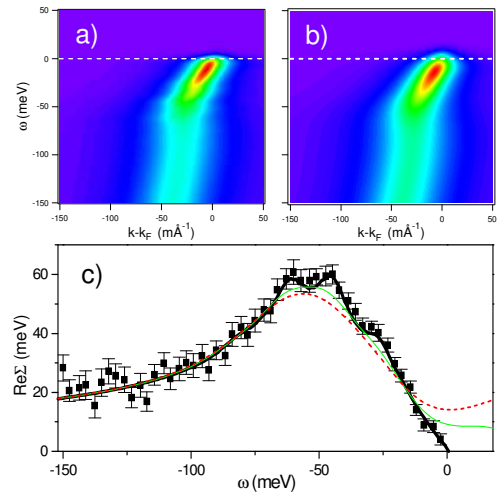} 
\end{center}
\caption{\label{fig1}
(a) The unbroadened spectral intensity at T=20 K simulated by using  
self-energy and bare dispersion from \cite{Zhou}. 
(b) Spectrum from (a), convoluted in energy by $\Delta_{exp}=12$ meV.
(c) The experimental (symbols) and calculated Re$\Sigma(\omega)$ from \cite{Zhou} (thick solid line). Also shown are the MDC extracted Re$\Sigma$ 
from simulated spectra when broadened with 12 meV (shown in (b)) (thin solid line) and 18 meV (dashed line) of the experimental energy resolution.}
\end{figure} 
In the following we completely neglect the momentum broadening and only convolve 
the simulated spectrum in energy. The result for $\Delta_{exp}=12$ meV is shown in Fig 1b, while the corresponding
MDC deduced Re$\Sigma(\omega)$ is shown in Fig 1c) by the thin solid line. If measured with $\Delta_{exp}=18$ meV, the Re$\Sigma(\omega)$ 
would look as the dashed line in Fig 1c).
It is obvious that the "fine structure", present in the unbroadened spectrum, dissappears completely, even at $\Delta_{exp}=12$ meV \cite{structure}. 
Therefore, its "presence" in the data signals some experimental error, most likely statistical noise. 
The finite energy resolution also introduces a shift in the apparent $k_F$ (an offset in Re$\Sigma(\omega)$ at $\omega=0$), 
altering the magnitude and the shape of the measured Re$\Sigma(\omega)$.
Depending on temperature and on details of quasiparticle dispersions, Fermi velocities may be significantly overestimated.
At low temperature ($kT\ll\Delta_{exp}$) the effect is most pronounced and near $\omega=0$ it is roughly proportional to $\Delta_{exp}/v_{F}$. 
In the opposite limit, ($kT\gg\Delta_{exp}$) these effects disappear. In the present case, however, the effect near $k_F$ is 
15-25$\%$ of the maximum value of Re$\Sigma(\omega)$ and should be easily detected. Even the 
shape and magnitude of the broad structure in Re$\Sigma(\omega)$ is significantly altered. However, Zhou \textit{et al} have 
missed and/or ignored these effects completely. Instead, they claim that much finer structure, unobservable under the cited experimental conditions, is real.
In conclusion, we showed that such structure can not be real and is probably noise related. 
\\
T. Valla\\
Physics Department, Brookhaven National Laboratory\\
Upton, NY 11973-5000\\


\begin{thebibliography}{99}
\bibitem{res} Fig. 1b) in \cite{Zhou} suggests that the actual energy resolution was $\sim22$ meV in that case.
\bibitem{structure} The simulated Re$\Sigma(\omega)$ has only one broad maximum at $\sim55$meV, while its second derivative obviously has more structure. 
The structure in the second derivative of the simulated curve can not be a valid criterion for the existence of multiple modes, especially in the case where 
the simulated curve differs so much from the data.
\end{thebibliography}
\end{document}